\documentclass[twocolumn,prb,aps,amsmath,showpacs,superscriptaddress,floatfix]{revtex4}
\usepackage{graphicx}
\usepackage{xspace} 

\newcommand{\tc}{\ensuremath{T_\mathrm{c}}\xspace}
\newcommand{\jc}{\ensuremath{J_\mathrm{c}}\xspace}
\newcommand{\ic}{\ensuremath{I_\mathrm{c}}\xspace}

\newcommand{\Jc}{\jc{}}

\newcommand{\dg}{\ensuremath{^\circ}\xspace}

\renewcommand{\vec}{\mathbf}

\begin{document}

\title{Imaging ac losses in superconducting films via scanning Hall probe microscopy}
\author{Rafael B. Dinner}
\author{Kathryn A. Moler}
\email[Electronic address: ]{kmoler@stanford.edu}
\affiliation{Geballe Laboratory for Advanced Materials, Stanford University, Stanford, California 94305}
\author{D. Matthew Feldmann}
\affiliation{Los Alamos National Laboratory, Los Alamos, New Mexico 87545}
\author{M. R. Beasley}
\affiliation{Geballe Laboratory for Advanced Materials, Stanford University, Stanford, California 94305}

\date{\today}

\begin{abstract}
Various local probes have been applied to understanding current flow through superconducting films, which are often surprisingly inhomogeneous. Here we show that magnetic imaging allows quantitative reconstruction of both current density, $J$, and electric field, $E$, resolved in time and space, in a film carrying subcritical ac current. Current reconstruction entails inversion of the Biot-Savart law, while electric fields are reconstructed using Faraday's law. We describe the corresponding numerical procedures, largely adapting existing work to the case of a strip carrying ac current, but including new methods of obtaining the complete electric field from the inductive portion determined by Faraday's law. We also delineate the physical requirements behind the mathematical transformations. We then apply the procedures to images of a strip of YBa$_2$Cu$_3$O$_{7-\delta}$ carrying an ac current at 400~Hz. Our scanning Hall probe microscope produces a time-series of magnetic images of the strip with 1~$\mu$m spatial resolution and 25~$\mu$s time resolution. Combining the reconstructed $J$ and $E$, we obtain a complete characterization including local critical current density, $E$--$J$ curves, and power losses. This analysis has a range of applications from fundamental studies of vortex dynamics to practical coated conductor development.
\end{abstract}


\maketitle

\section{\label{intro}Introduction}

\begin{figure}[tb]
  \includegraphics[height=4.5in]{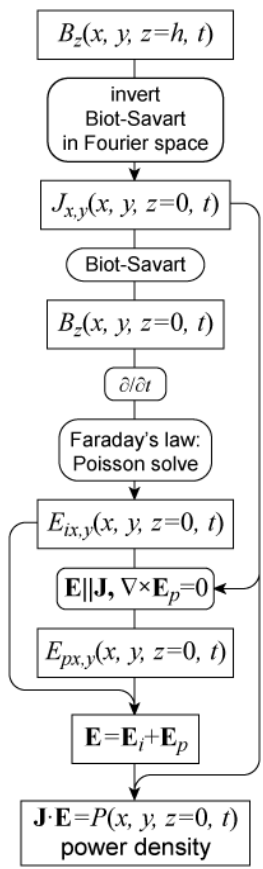}
  \caption{\label{fig:flow}A flow chart illustrating how we use a series of magnetic field images to reconstruct current density and electric field in the sample, and finally power input.}
\end{figure}

Developing high-temperature superconducting films to carry high current densities with low dissipation has proved to be an enormous challenge, resulting in composite materials that are inhomogeneous down to the atomic scale.\cite{cc} For characterizing inhomogeneous current flow, magnetic imaging has emerged as an important tool. By measuring the magnetic field in a plane above a film's surface, one can invert the Biot-Savart law to reconstruct a two-dimensional map of current density, $\vec{J}$, in the film, as described in Section~\ref{j_method}.\cite{wikswo,cg,reg} Such a measurement can be obtained by various methods including magneto-optical imaging,\cite{feldmann_mo_cr,jooss_review,wijngaarden_mo_review}, scanning Hall probe microscopy\cite{rsi,shp_mo_cr}, and scanning superconducting quantum interference device (SQUID) microscopy.\cite{kiss_squid_slm,ssquid}

Other imaging techniques, such as scanning potentiometry,\cite{potentiometry_on_cc} scanning laser microscopy,\cite{kiss_squid_slm,ltslm} and scanning electron microscopy\cite{ltsem} can map out the electric field, $\vec{E}$, that arises from vortex movement or other changes in local supercurrent density. In these techniques, to generate measurable electric fields, the superconductor is biased with a current $I$ slightly greater than the dc critical current \ic at which vortices start to flow. In principle, such measurements could be combined with magnetic imaging of the same sample, and together, $\vec{J}$ and $\vec{E}$ would provide a complete, spatially-resolved electrical characterization of the material, including the local critical current density \jc, and the local power input, which can be calculated as $\vec{J}\cdot\vec{E}$.

In this work, we demonstrate that for $I<\ic$, time-resolved magnetic imaging can simultaneously determine both $\vec{J}$ and $\vec{E}$ in a superconducting film. Figure~\ref{fig:flow} summarizes the operations involved. The instantaneous magnetic field $\vec{B}$ determines $\vec{J}$, while the time rate of change of $\vec{B}$ is related to $\vec{E}$ through Faraday's law,
\begin{equation}
\vec{\nabla}\times\vec{E}=-\partial _t \vec{B}.
\end{equation}
This relation only constrains the inductive portion of electric field, $\vec{E}_i$. To reconstruct the remaining electrostatic part, $\vec{E}_p$, we must impose the additional restrictions that $I<\ic$, as discussed in Section~\ref{e}, and that $\vec{E}$ is parallel to $\vec{J}$, discussed in Section~\ref{ep}. However, many important applications that are not accessible to techniques operating above \ic do lie within these restrictions, such as ac losses in superconducting films.

Faraday's law has been applied previously to derive ac loss from a critical state model of magnetic fields in a homogeneous, infinitely long superconducting wire.\cite{norris,brandt} In the present work, this method is reformulated to allow for the inhomogeneity of a real conductor revealed by our magnetic images. The method also applies to the case of magnetization decay due to flux creep, which was recently analyzed using magneto-optic images.\cite{jooss}

Here, we use a cryogenic scanning Hall probe microscope (SHPM) to obtain a series of images of $B_z$, the component of magnetic field perpendicular to the sample surface, as it evolves with time. The microscope is described in Ref.~\onlinecite{rsi} and further information is given in Section~\ref{apparatus}.

To analyze ac losses, we image a strip of the high-temperature superconductor YBa$_2$Cu$_3$O$_{7-\delta}$ (YBCO), described in Section~\ref{sample}. We apply an ac current to the strip at 400~Hz, a typical operating frequency for applications,\cite{400Hz} and must image the magnetic response faster than this. Our scan speed---a few pixels per second---is slow in comparison, so instead of acquiring an entire sequence of images within one cycle of applied current, we obtain an average response over many cycles, as explained in Section~\ref{daq}, with 25~$\mu$s time resolution.

Section~\ref{b} presents this time series of magnetic images, which are then transformed into images of current and electric field in Sections~\ref{j} and \ref{e}. These quantities are combined in Sections~\ref{power} and \ref{ej} to yield maps of dissipation and superconducting characteristics.

\section{\label{experiment}Experimental procedure}

\subsection{\label{apparatus}Apparatus: scanning Hall probe microscope}
The apparatus is described in Ref.~\onlinecite{rsi}, but essential experimental details and notable modifications are included here. The instrument generates magnetic images by rastering a Hall sensor above the sample surface, measuring magnetic field at each pixel. The sensor's Hall resistance is approximately proportional to $B_z$, the component of magnetic field perpendicular to the sample surface, with a field resolution of $4\times10^{-3}$~G/$\sqrt{\text{Hz}}$.

The scanning stage, based on stepper motors driving micrometers, offers a $1\times4$~cm scan area for aligning to samples and macroscopic features, and can zoom in for 200~nm positioning resolution. However, the image resolution is limited by the sensor: First, its lithographic size leads to averaging of the field over 500~nm. Second, though its tip remains in contact with the sample during scanning, the sensor's slight tilt lifts the sensitive area 1~$\mu$m above the sample surface. These factors yield 1~$\mu$m spatial resolution.

The sensitive area is coated with a grounded gold gate to screen electric fields, followed by an insulating aluminum oxide layer to isolate the gate from sample voltages.\cite{ald} The oxide is also intended to provide protection against mechanical wear.

\subsection{\label{sample}Sample: YBCO strip}
The YBCO film studied is 180~nm thick, grown epitaxially by pulsed laser deposition (PLD) on a SrTiO$_3$ [001] substrate. Photolithographic patterning followed by argon ion milling removes parts of the film, leaving a bridge as shown in Fig.~\ref{bjep}(a). The substrate is held in vacuum, attached to the microscope's copper coldfinger by a thin layer of low-temperature varnish. The coldfinger is cooled by a continuous flow of liquid helium. The film's \tc (defined by the maximum in $dR/dT$) is measured in this cryostat to be 90~K. For imaging, the film's coldfinger is held at 40~K while the Hall sensor is held at 54~K. However, the current return lead (to the right of the segment imaged) is narrower than the bridge (32~$\mu$m versus 50~$\mu$m), and a magnetic scan (not shown) indicates that the applied current of 0.75~Arms exceeds its critical current. The consequent dissipation in the return lead may heat the sample several degrees above 40~K.

\subsection{\label{daq}Data acquisition}
The images shown in Fig.~\ref{bjep}(c)--(j) are all derived from one scan as described in Ref.~\onlinecite{rsi}. A 402.7~Hz ac current is applied continuously to the sample. The sensor rasters over a 100 by 150~$\mu$m area with 0.5~$\mu$m pixel spacing. It pauses at each pixel and records the waveforms $I(t)$ and $B(t)$ for approximately 80 cycles of applied current. These cycles are overlaid and averaged. Values from the averaged waveforms are collated into images by their phase within the cycle. The complete set of images is presented as the movie \texttt{BJE.avi} in the supplemental material.\cite{mgroup_web} The images correspond to 100 time slices over the cycle of applied current, each representing the average magnetic field over a 25~$\mu$s interval.

\section{\label{b}Magnetic image results}

\begin{figure*}[tb]
  \includegraphics[width=\textwidth]{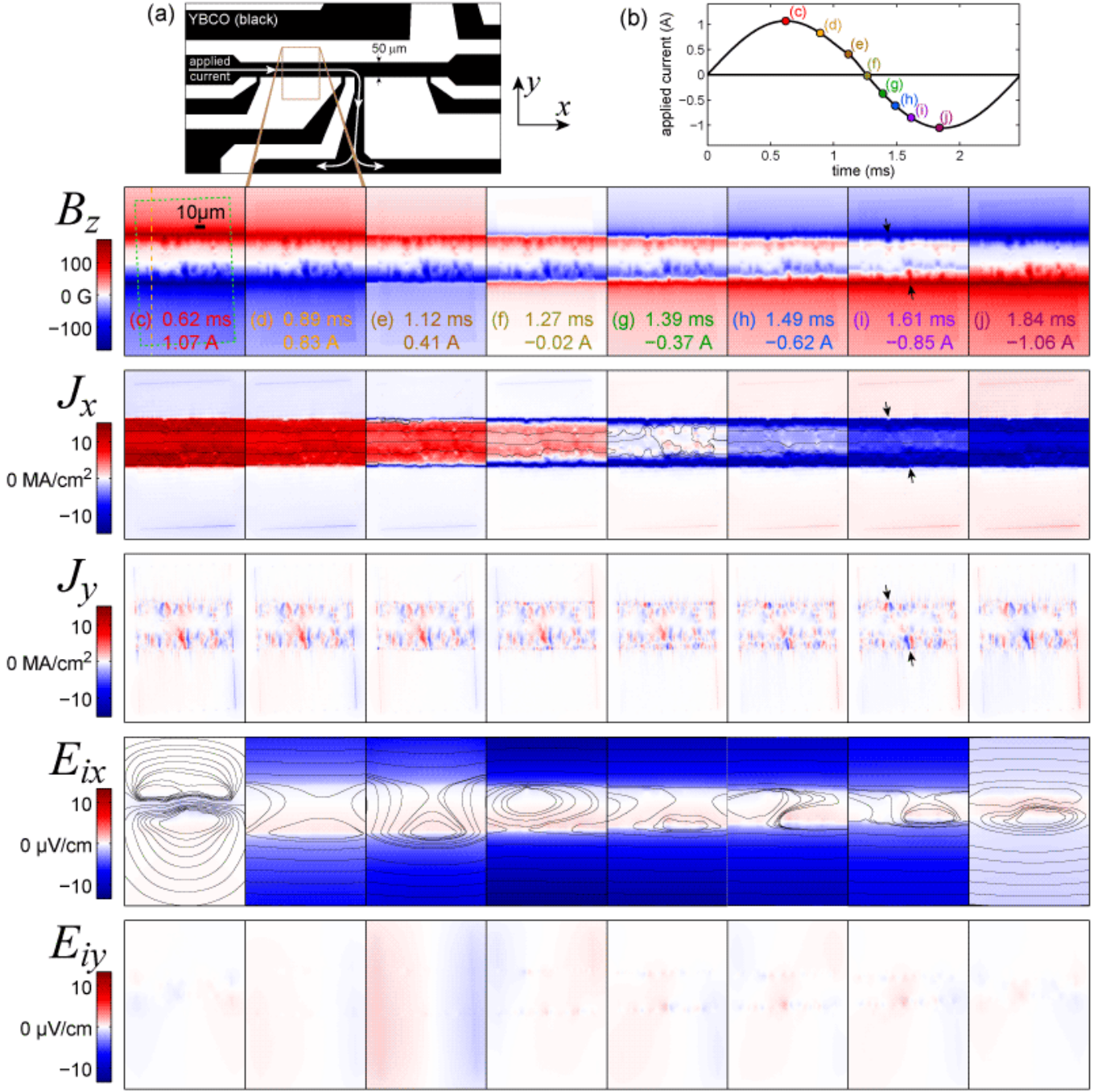}
  \caption{\label{bjep}(Color online) Imaging current-induced flux penetration into a YBCO film: (a)~Illustration of the sample geometry. The approximate area of the magnetic images is outlined by the brown box. Current is injected at the left and extracted from the downward-facing lead; the right-facing current lead and smaller voltage leads are not used (floating). (b)~Applied current during a 0.75 Arms, 400~Hz cycle. Select times are marked, and the corresponding magnetic images are shown in (c)--(j). Image~(c) indicates the scale and boundary of the experimental data (dashed green box); the background outside of the data is filled by a fit to a critical state model. The vertical dashed yellow line indicates the location of the cross sections shown in Fig.~\ref{fig:cut}. In image~(i), the black arrows point out two spots where vortices enter the film more easily. Below (c)--(j) are reconstructions from the magnetic data: components $J_x$ and $J_y$ of the current density flowing in the sample, and components $E_{ix}$ and $E_{iy}$ of the inductive portion of the electric field. Color scales for $x$ and $y$ components are the same. Black streamlines of $\vec{J}$ and $\vec{E}_i$ overlay their $x$ components. The complete set of frames is presented as the movie \texttt{BJE.avi} in the supplemental material.\protect\cite{mgroup_web}}
\end{figure*}

\begin{figure}[tb]
  \includegraphics[width=8.6cm]{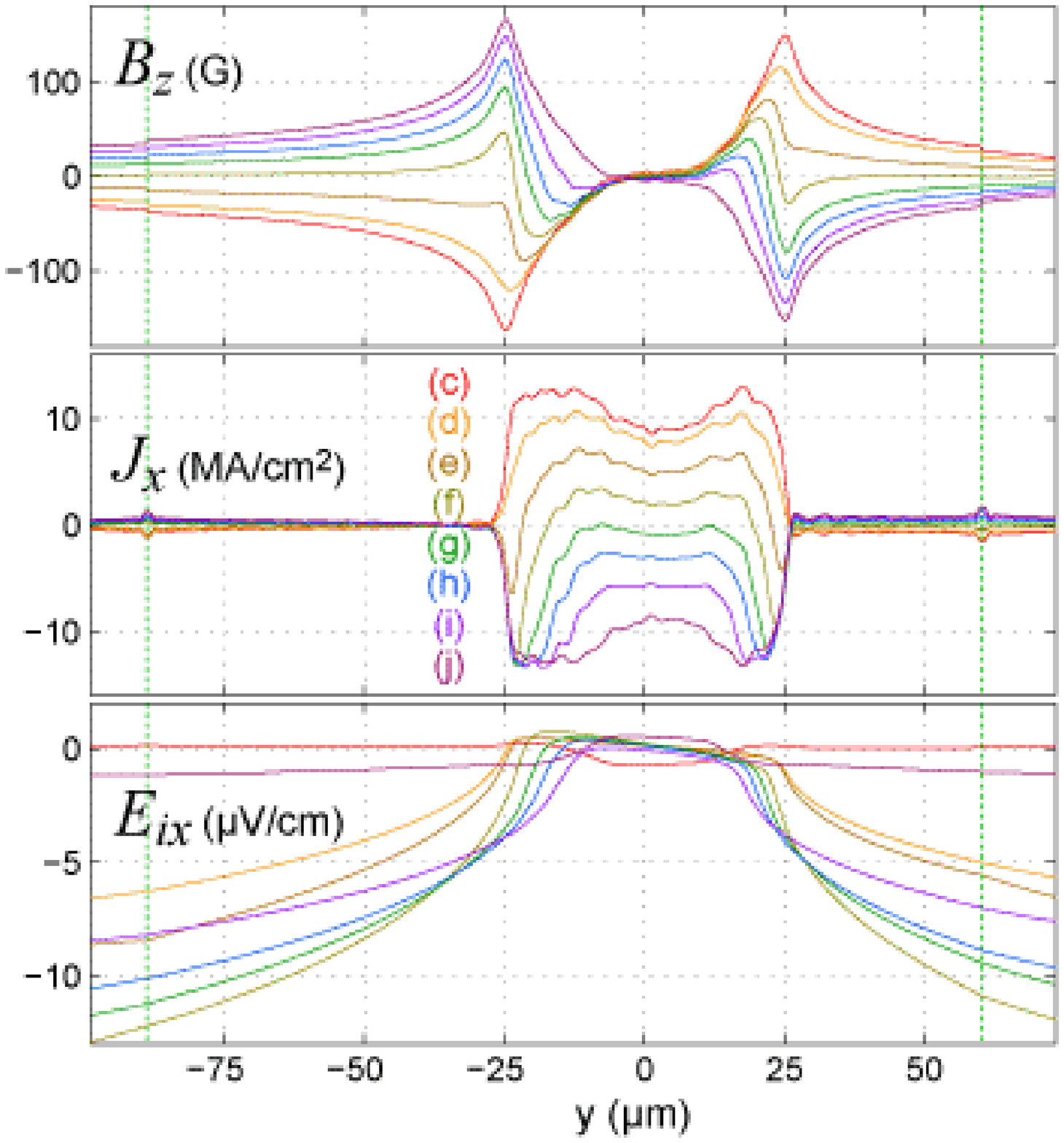}
  \caption{\label{fig:cut}(Color online) Cross sections of the data in Fig.~\ref{bjep}. The sections are taken along the $y$ axis at the location of the dashed yellow line in Fig.~\ref{bjep}(c). The vertical dashed green lines mark the boundary between data and background fit. Successive frames in the current cycle are overlaid; their colors and labels match the frame labels in Fig.~\ref{bjep}.}
\end{figure}

This work is primarily concerned with deriving quantities of interest from time-resolved magnetic images, but it is worth first examining the images directly. From them, we extract a great deal of qualitative information that shapes the assumptions under which further quantities are derived.

Figure~\ref{bjep}(c)--(j) are selected from a larger set of frames, and represent one half-cycle of ac applied current. Initially, the maximum current $I$=1.07 A is applied and flux has penetrated, somewhat inhomogeneously, into both edges of the strip. $I$ is apparently below the critical current, $I_c$, as a central flux-free region separates the two flux fronts. This condition is necessary for our calculation of electric field, as explained in Section~\ref{e}.

By (f), the applied current is reduced to zero, and vortices remain trapped in the edge regions. Flux of the opposite sign starts to enter at the edges. We then see this opposite flux erase and replace the trapped flux as negative current is applied. The succeeding half of the cycle approximately repeats the $B$, $J$, and $E$ configurations shown, but with opposite signs.

The boundary of the data is shown as a dashed green box in (c). The surrounding background is calculated with a critical state model for a thin superconducting strip.\cite{brandt,strip_model} The main purpose of this background is to better match the boundaries of the image for the Fourier transformations described in Section~\ref{j_method}. The good agreement between the model and the data also shows that much of the strip's response can be ascribed to critical state behavior, though deviations, such as spatial inhomogeneity in pinning strength, are evident.

The model is fit to the entire set of frames at once. The free parameters are the height of the sensor above the sample plane, 1.2~$\mu$m, the sensor tilt, 3\dg about the $x$ axis, the strip's critical current, 1.13~A, the Hall coefficient, 0.10~$\Omega$/G, and the amplitude, 7.2~Oe~rms, of an applied field proportional to the applied current, explained below. The Hall coefficient's fit value is consistent with the calibrated value of 0.11~$\Omega$/G. The $y$ position and tilt of the strip about the $z$ axis are also allowed to vary, and the tilt is zeroed by rotating the image. These parameters are constrained to be constant over time; the only change in the calculation from frame to frame is the (known) applied current.

Adding the small, uniform applied field that varies with the applied current improves the fit, and is suggested by the sample geometry in Fig.~\ref{bjep}(a), where the current returns to the right and below the bridge. Positive returning current generates a negative field at the section of bridge imaged, which accords with the sign of the field added to the calculation. Furthermore, the vertical segment of the return lead should add a negative $dB/dx$, which is not accounted for by the calculation, and which does explain why the disparity between the data and the calculation is largest toward the lower right corner of the boundary.

\section{\label{j}Current reconstruction}

\subsection{\label{j_method}Magnetic inversion with regularization}

The Biot-Savart law describes the magnetic field $\vec{B}$ generated by a current distribution $\vec{J}$,
\begin{equation}
\vec{B}(\vec{r}) = \frac{\mu_0}{4\pi} \int d^3\vec{r'} \frac{\vec{J}(\vec{r'})\times(\vec{r}-\vec{r'})}{\left|\vec{r}-\vec{r'}\right|^3}.
\label{biot-savart}
\end{equation}
Several authors have tackled the problem of inverting this relation to obtain a planar current distribution $\vec{J}(x,y)$ from a planar magnetic measurement $B_z(x,y)$.\cite{wikswo,cg,reg} We tried two of these existing methods: regularization\cite{reg} and conjugate-gradient.\cite{cg} The conjugate-gradient method produces current distributions dominated by unphysical artifacts. Further testing with simulated data suggests that this occurs when the current distribution extends outside the image boundaries, as in our images, which encompass only a section of the superconducting strip.

We met with greater success using the regularization method with generalized cross-validation (GCV), described in detail in Ref.~\onlinecite{reg} and summarized here. Taking the $z$ component of Eq.~\ref{biot-savart} and Fourier transforming in $x$ and $y$, we find an algebraic relation between the Fourier transformed quantities $\tilde{B_z}$, $\tilde{J_x}$, and $\tilde{J_y}$:
\begin{equation}
\tilde{B_z}(k_x, k_y, z) = \frac{\mu_0}{2} e^{-kz} \frac{i}{k} \left( k_y \tilde{J_x} - k_x \tilde{J_y} \right).
\label{bj_eq}
\end{equation}
We assume that $J_z$, current flowing perpendicular to the plane of the film, is insignificant. We also ignore any $z$ dependence of the in-plane components of current, approximately solving for the current density averaged over the film thickness. These are reasonable approximations in our film's geometry; its thickness, 180~nm, is smaller than its effective London penetration depth, $\lambda_{\mathrm{eff}} = \lambda_{ab}/\tanh(d/2\lambda_{ab}) = 400$~nm.\cite{ybco_penetration_depth,vortex_thick_film} The specific kernel that we use to relate $J_x$ and $J_y$ to $B_z$ is that of a film of 180~nm thickness.\cite{reg}

We now reduce $J_x$ and $J_y$ to a single unknown by noting that $\vec{J}$ must be nearly divergence-free at the operating frequency of 400~Hz (the resonant frequency, $1/\sqrt{LC}$, of a piece of strip like that imaged would be $\sim100$~GHz). This allows us to derive both $J_x$ and $J_y$ from the local magnetization $g(x,y)$,\cite{brandt_prl} where
\begin{equation}
\vec{J}(x,y)=-\hat{z}\times\vec{\nabla}g(x,y) \Rightarrow J_x = \partial _y g \mbox{, } J_y = -\partial _x g.
\label{jg_eq}
\end{equation}
In Fourier space,
\begin{equation}
\tilde{g}(k_x,k_y)=\frac{i}{k^2} \left( k_y \tilde{J_x} - k_x \tilde{J_y} \right).
\label{gj_eq}
\end{equation}
Eqs.~\ref{bj_eq} and \ref{gj_eq} yield
\begin{equation}
\tilde{g} =  \frac{2}{\mu_0} e^{kz} \frac{1}{k} \tilde{B_z}.
\label{gb_eq}
\end{equation}
Thus our basic procedure is to Fourier transform $B_z$, solve for $\tilde{g}$, Fourier transform back to the real space $g(x,y)$, and use Eq.~\ref{jg_eq} to obtain $J_x$ and  $J_y$.

The first difficulty arises from the factor $e^{kz}$ in Eq.~\ref{gb_eq}. Spurious high spatial frequencies (with wavelength greater than $z$, the measurement height) in the magnetic data are exponentially amplified in the inverted current. While scanning Hall probe microscopy enjoys lower noise and smaller $z$ than magneto-optical imaging,\cite{bending_review} noise in our images can still dominate the reconstructed $J$. The method of regularization compensates by suppressing high frequencies (smoothing), and GCV determines an optimal amount of regularization from the data itself. In practice, we used GCV as a guide to choose a regularization parameter ($\lambda=100$ as described in Ref.~\onlinecite{reg}) that we held fixed across the set of frames.

Similarly, differentiating $g$ to find the components of $J$ amplifies high-frequency noise, so we use Savitsky-Golay smoothing to extract the derivatives. The smoothing is quadratic with a frame size of 2.5~$\mu$m (5 pixels).

The second difficulty is that the Fourier transform of $B$ assumes periodicity in the vertical and horizontal directions, and mismatches between the left and right, and top and bottom boundaries of the $B$ image lead to artifacts dominating the image once it has been transformed, manipulated, and transformed back. A common solution is to window the data, bringing its boundaries smoothly to zero. This discards a large portion of each image, however.

Instead, we rotate the original data (2.3$^\circ$ about the $z$ axis) so that the bridge runs horizontally, then center it on a larger area (a square 256 $\mu$m on a side) in order to move the edge effects away from the data. We fill in the background with a calculated $B$ as described in Section~\ref{b}. Only after calculating $\vec{J}$ and $\vec{E}_i$ (described in Section~\ref{ei}) do we crop the images back to the original dimensions (plus the margin seen in Fig.~\ref{bjep}(c)). Finally, before Fourier transforming, we mirror the images top-to-bottom in order to better match the top and bottom edges without going to the much larger area necessary to allow the field to die off, which would be more computationally cumbersome.

While these preparations remove artifacts associated with edge mismatch, we do observe a spurious bump in the reconstructed current at the boundary between data and fit, seen clearly in the cross section of $J_x$ in Fig.~\ref{fig:cut}. However, this artifact is about ten times smaller than our signal, and appears to be confined to the boundary.

\subsection{\label{j_discussion}Discussion of current density images}

The results largely agree with our expectations for a superconducting strip. The reconstructed current flows within the strip approximately in the $x$ direction. At the maximum applied current (Fig.~\ref{bjep}(c)), the current density $\left(J=\sqrt{J_x^2+J_y^2}\right)$ in the edge regions of flux penetration should equal the critical current density, according to the critical state model. $J$ then dips down (but is not zero) in the vortex-free central region, as expected from the demagnetization effect of the strip geometry.\cite{brandt,strip_model} This separation into edge and central regions is clearest in the cross section of $J_x$, Fig.~\ref{fig:cut}(c). $J$ averaged over the edge regions is 12 MA/cm$^2$, which accords with macroscopic transport measurements of \jc on a similar film (9 MA/cm$^2$ at 44~K).\cite{daniels_jc} We expect transport, which is sensitive to the weakest point in a superconductor, to yield a lower value than the spatially averaged \jc.

The cross section of $J_x$ also shows more current on the $-y$ side, indicative of the applied field (described in Section~\ref{b}) modifying the symmetric distribution one would expect for an applied current. While the cross sections vary along the length of the strip, this asymmetry is typical.

$J_y$, plotted on the same color scale as $J_x$, is smaller, but highlights where the current reroutes around apparent weak spots in the film. Spots producing the largest $J_y$ are marked with arrows in Fig.~\ref{bjep}(i). These features in $J_y$ correspond to bumps in the streamlines overlaid on $J_x$ where the current spreads around these defects. In the following Sections, we show that these are also spots of high electric field and dissipation.

\section{\label{e}Electric field reconstruction}
We relate electric field to the magnetic field we measure via Faraday's law,
\begin{equation}
\vec{\nabla}\times\vec{E}=-\partial _t \vec{B}.
\end{equation}
Taking the $z$ component,
\begin{equation}\label{ecurl}
\partial _x E_y - \partial _y E_x = -\partial _t B_z.
\end{equation}
This only defines $\vec{E}$ up to the gradient of a scalar. We therefore use the Helmholtz decomposition to separate the electric field into a divergence-free inductive portion, $\vec{E}_i$, and a curl-free electrostatic portion, $\vec{E}_p$ (following the notation of Ref.~\onlinecite{jooss}),
\begin{equation}
\vec{E} = \vec{E}_i + \vec{E}_p.
\end{equation}

$\vec{E}_p = -\vec{\nabla} \phi$ where $\phi$ is a scalar potential and $\vec{\nabla}\cdot\vec{E}_p = \rho/\epsilon_0$, the charge density. Our measurements determine $\vec{E}_i$ through Eq.~\ref{ecurl}, but do not determine $\vec{E}_p$ without further constraints, as described in Section~\ref{ep}.

To illustrate: If we apply a dc current $I>\ic$ to our strip, flux flow or other resistive behavior generates an $\vec{E}_p$, but we would see little time variation of the magnetic field when averaging over length scales greater than the intervortex spacing. In contrast, for $I<\ic$, the voltage and electric field (both $\vec{E}_i$ and $\vec{E}_p$) are zero in the steady state. We therefore remain below \ic of our superconducting strip in order to minimize unmeasurable portions of $\vec{E}_p$.

Furthermore, below \ic, even in a dynamic state, $\vec{E}_p$ remains zero for a strip that is uniform in $x$ with no Hall effect. In this case, symmetry dictates that $E_y = 0$, $E_z = 0$, and $\partial_x E_x = 0$, thus $\vec{\nabla} \cdot \vec{E} = 0$. Then $\vec{E}_p$ is uniform, and zero below \ic. Thus all of the behavior we expect from a model strip will be contained in $\vec{E}_i$.

\subsection{\label{ei}Inductive electric field $\vec{E}_i$}

To reconstruct the divergence-free $\vec{E}_i$, we proceed as for $\vec{J}$, solving for a potential function $h$ where
\begin{equation}
\vec{E}_i(x,y)=-z\times\vec{\nabla}h(x,y) \Rightarrow E_{ix} = \partial _y h \mbox{, } E_{iy} = -\partial _x h
\end{equation}
in which case Eq.~\ref{ecurl} becomes a Poisson equation for $h$:
\begin{eqnarray}
-\partial^{2}_{x}h - \partial^{2}_{y}h & = & -\partial _t B_z \\
\label{poisson} \nabla ^{2}_{2D} h & = & \partial _t B_z
\end{eqnarray}

We are interested in the electric field in the sample plane ($z=0$) rather than the measurement plane. We therefore need $B_z(z=0)$, which we obtain from the current distribution via the Biot-Savart law. $B_z$ at any height is easily computed by rearranging Eq.~\ref{gb_eq} to obtain $\tilde{B_z}$ in terms of $\tilde{g}$, the magnetization function from which $\vec{J}$ is derived:
\begin{equation}
\tilde{B_z} = \frac{\mu_0}{2} e^{-kz} k \tilde{g}.
\label{bg_eq}
\end{equation}

It is also easiest to solve the Poisson equation in Fourier space, where Eq.~\ref{poisson} becomes
\begin{equation}
(i k)^2 \tilde{h} = \partial _t \tilde{B}_z(z=0)
\label{poisson_fft_eq}
\end{equation}
and, combining Eqs.~\ref{bg_eq} and \ref{poisson_fft_eq},
\begin{equation}
\tilde{h} = -\frac{\mu_0}{2k} \partial _t \tilde{g}
\end{equation}

To approximate the time derivative of $\tilde{g}$, we compute twice the final number of frames (200 frames spaced by 12.5~$\mu$s, but each still averages a 25~$\mu$s interval) and take the differences between successive even frames. We can then calculate simultaneous electric field and current at the times of the odd frames.

Once we have solved for $\tilde{h}$ and transformed back to real space, we use Savitsky-Golay smoothing, as with $J$, to extract the partial derivatives corresponding to $E_{ix}$ and $E_{iy}$.

Finally, Eq.~\ref{ecurl} only defines $E_{ix}$ and $E_{iy}$ up to constants. Setting the constants is equivalent to finding a field-free point, to which Norris devoted much care.\cite{norris} We set the zeros based on the edges of the uncropped images, as far from inhomogeneities in the data as possible. For the $E_{ix}$ zero we use the mean of the two pixels at the vertical center of the strip (where we expect no $E$, as Norris pointed out) on the left and right edges. We set the $E_{iy}$ zero to the mean of the four corners. This mean is zero for an ideal strip, and minimally affected by fields originating toward the center of the image.

\subsection{\label{ei_discussion}Discussion of $\vec{E}_i$ images}
As the applied current decreases over the half cycle shown, $E_{ix}$ remains approximately zero (white) in a central region of the strip. Outside of this region it becomes negative (blue), continuing past the strip edges. This is also visible in the cross sections of $E_{ix}$ in Fig.~\ref{fig:cut}. This behavior accords with our expectations for vortices moving into the edges of the film. The central region shrinks, tracking the flux front as vortices enter. Although vortices and current from the previous half cycle are present inside the central region, the vortices remain pinned and therefore do not generate an electric field.

The central region is not completely field-free, however. Interestingly, the field it does display---about 10 times smaller than the edge fields---is maximal and opposite to the current when $dI/dt=0$, a point where our critical state model would dictate $E=0$. Such a field would arise, though, from a relaxation of \jc, i.e. flux creep in which the vortices continue to move into the strip even as the current momentarily stops ramping. It leads to a (temporary) negative power input to the film, discussed in Section~\ref{power}.

Such movement while $dI/dt=0$ is also visible in the full set of magnetic images. While the magnetic field evidence alone is subject to errors in phase relative to the applied current, the electric field confirms that the relaxation is real. Such relaxation is the focus of Ref.~\onlinecite{jooss}.

Finally, the cross sections of $E_{ix}$ reveal an unexpected negative tilt, $dE/dy$, both in the central and outer regions, which is an error that arises from the tilt of the Hall sensor, as shown in Section~\ref{err}.

\subsection{\label{ep}Electrostatic electric field $\vec{E}_p$}

We established in Section~\ref{e} that for a uniform superconducting strip with no Hall effect below \ic, $\vec{E}_p=0 \Rightarrow \vec{E}=\vec{E}_i$, even with an ac current. Indeed, the $\vec{E}_i$ that we observe contains all the features we expect from a uniform strip, as discussed in Section~\ref{ei_discussion}. Our images, however, also reveal inhomogeneity, which could produce a non-zero $\vec{E}_p$. 

Here we show that in spite of this inhomogeneity, we expect the total electric field $\vec{E}$ to remain approximately parallel to $\vec{J}$ locally, which proves sufficient to reconstruct $\vec{E}_p$ as the field that compensates for any component of $\vec{E}_i$ perpendicular to $\vec{J}$. Our method is inspired by Ref.~\onlinecite{jooss}, but we do not make the additional and incorrect assumption that the component of $\vec{E}_p$ parallel to $\vec{J}$ is zero.\cite{jooss_incorrect} We also describe a rather different route to $\vec{E}$ and $\vec{E}_p$ in Section~\ref{curlej}.

\subsubsection{Validity of the constraint $\vec{E}\|\vec{J}$}

First we must justify that in our experiment, $\vec{E}$ is parallel to $\vec{J}$, emphasizing that this will not be true for all materials. For example, a material can have an intrinsic Hall effect. However, macroscopic transport measurements indicate that for YBCO in the superconducting state, the Hall effect is insignificant; the component of $\vec{E}$ perpendicular to $\vec{J}$ is at most 1000 times smaller than the parallel component.\cite{ybco_hall}

Another violation of $\vec{E}\|\vec{J}$ could arise from a feature such as a grain boundary, whose orientation prevents vortices from moving perpendicular to $\vec{J}$. More generally, any gradient in superfluid condensate energy density will exert a force on vortices. We show here, however, that at least on lengths scales greater than the image resolution of 1~$\mu$m, we can rule out the presence of gradients strong enough to compete with the pinning forces that occur on the scale of the coherence length, $\xi \approx 2$~nm. We put an upper bound on such a gradient in our material by assuming that the gradients in pinning strength that we observe stem entirely from changes in condensate energy density. We take $J$ (the magnitude of the current density) at maximum applied current as a map of \jc near the edges of the strip (as discussed in Section~\ref{j_discussion}). The magnitude of the pinning force per length for a single vortex is
\begin{equation}
F_p = \Phi_0 J_c
\end{equation}
from which we estimate the depth of the vortex pinning potential as $V_p = F_p\xi$. The line energy of the vortex will be $\sim V_p \cdot 10$, from the ratio of the calculated depairing current to the measured \jc.\cite{gurevich_talk} Thus the gradient of $F_p$ yields a gradient of line energy, i.e. a force $F_{grad}$, whose magnitude we compare to $F_p$. We find that $F_p/F_{grad}$ averages 2000 over the edge regions of the strip with a minimum of 200.

This analysis suggests that in our material, the Lorentz force from the current at \jc, along with pinning forces, dominate other forces felt by vortices, so that vortices move perpendicular to $\vec{J}$ and generate $\vec{E}$ parallel to $\vec{J}$. However, we do not claim to rule out every possible materials effect, e.g. effects that average out below our resolution of 1~$\mu$m.

\subsubsection{Method of reconstructing $\vec{E}_p$}

\begin{figure*}[tb]
  \includegraphics[width=17.8cm]{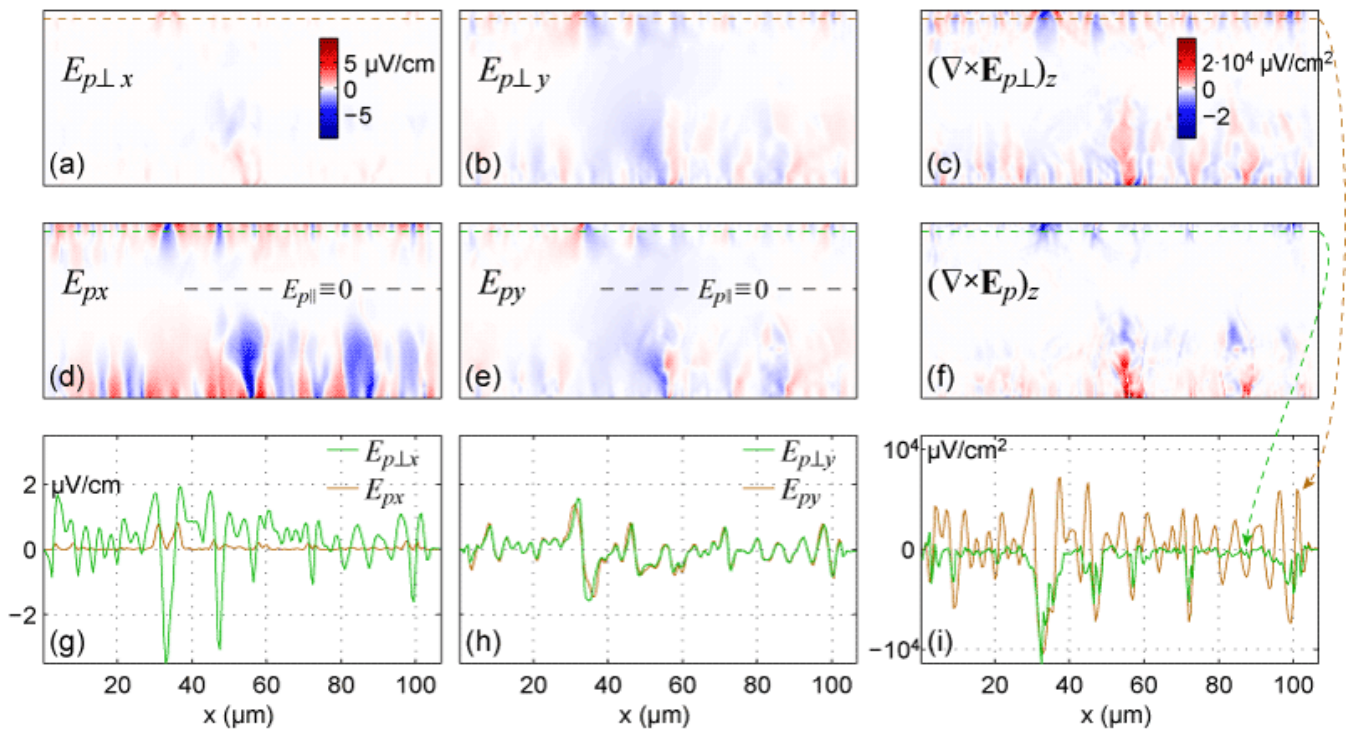}
  \caption{\label{ep_fig}(Color online) Reconstructing the electrostatic portion of electric field, $\vec{E}_p$: All images are cropped to the area of the strip. The electric fields are on the same color scale, shown in (a), and the curls are on the scale shown in (c). (g)--(i) are $x$ cross sections from the images above them, with locations indicated by the dashed lines of matching color. The cross sections also indicate the spatial scale of the images. The component of $E_p$ perpendicular to the current, (a)--(b), is obtained from $E_i$ under the assumption $\vec{E} \| \vec{J}$. However, (c) shows that this component alone does not satisfy $\vec{\nabla}\times\vec{E}_p=0$. We calculate and add a parallel component, yielding the complete $\vec{E}_p$ in (d)--(e). Our $E_{p\|}$ calculation requires an integration constant, defined by assuming $E_{p\|}=0$ along the lines shown in (d) and (e). Adding $E_{p\|}$ suppresses, but does not perfectly cancel, $\vec{\nabla}\times\vec{E}_p$, as seen in (f) compared to (c), and from the cross sections in (i).}
\end{figure*}

Starting with $\vec{E}_i$ and $\vec{J}$, reconstructed in Sections~\ref{ei} and \ref{j}, and the constraint that $\vec{E}=\vec{E}_i+\vec{E}_p$ is parallel to $\vec{J}$, we can solve for $\vec{E}_p$. The component of $\vec{E}_p$ perpendicular to $\vec{J}$ must cancel that of $\vec{E}_i$:
\begin{equation}
E_{p\bot} = -E_{i\bot}
\label{eiep}
\end{equation}
where $E_{i\bot}$ is calculated by subtracting from $\vec{E}_{i}$ its projection onto $\vec{J}$. Having obtained $E_{p\bot}$, we construct $E_{p\|}$, the component of $\vec{E}_p$ parallel to $\vec{J}$, to satisfy
\begin{equation}
\vec{\nabla} \times \vec{E}_p = \vec{0}
\label{curlep_eq}
\end{equation}
which follows from the definition of $\vec{E}_p$. The $z$ component of Eq.~\ref{curlep_eq} tells us that at each point,
\begin{equation}
\partial_\bot E_{p\|} = \partial_\| E_{p\bot}
\label{curlep}
\end{equation}
or more explicitly,
\begin{equation}
-\frac{J_y}{J} \partial_x E_{p\|} + \frac{J_x}{J} \partial_y E_{p\|} = \frac{J_x}{J} \partial_x E_{p\bot} + \frac{J_y}{J} \partial_y E_{p\bot},
\end{equation}
which is a linear, first-order partial differential equation for $E_{p\|}(x,y)$.

We tried two approaches to solving this equation for $E_{p\|}$. In the first, we start with a trial solution, $\vec{E}_{p1}$, composed of the known $E_{p\bot}$ and $E_{p\|}=0$. We calculate its curl $(\vec{\nabla}\times\vec{E}_p)_z$, which will be zero for the true $\vec{E}_p$. From the curl, we reconstruct a divergence-free field $\vec{F}$ following the same procedure we used to reconstruct $\vec{E}_i$ from $B_z$, described in Section~\ref{ei}. The reconstruction preserves $\vec{\nabla}\times\vec{F} = \vec{\nabla}\times\vec{E}_{p1}$, so $\vec{G} \equiv \vec{E}_{p1} - \vec{F}$ is curl-free, as desired. However, $\vec{G}$ does not preserve $E_{p\bot}$. So we construct a new trial $\vec{E}_{p1}$ consisting of the components $G_\|$ and $E_{p\bot}$, and iterate. The true $\vec{E}_p$ would remain unchanged by such a procedure. In practice, however, we found that the procedure misconverged to an $\vec{E}_{p1}$ with larger curl than the initial trial.

We therefore attempted a more direct integration of Eq.~\ref{curlep}:
\begin{equation}
E_{p\|} = \oint \partial_\| E_{p\bot} ds
\label{integral_ep}
\end{equation}
where the integral is taken along a path $\vec{s}$ that remains perpendicular to $\vec{J}$ at each point along its length, and starts at some $(x,y)$ such that $E_{p\|}(x,y)=0$.

The result, constructed from many such paths, is shown in Fig.~\ref{ep_fig}. The initial data ($\vec{J}$ and $\vec{E}_i$) is taken from frame h in Fig.~\ref{bjep}, then cropped to the area of the strip. For starting points, we set $E_{p\|}(x,y)=0$ along the horizontal line shown in Fig.~\ref{ep_fig}(d). In practice, to ensure coverage of every pixel, we start a path at each pixel and work back to the zero line. For efficiency, we skip pixels that have been covered by previous paths.

$E_{p\bot}$, shown in Fig.~\ref{ep_fig}(a)--(b), represents the starting data from which we calculate $\partial_\| E_{p\bot}$. Integration gives us $E_{p\|}$, which we add (as vector components) to $E_{p\bot}$ to obtain (d)--(e). The method is far from perfect, as evinced by the non-zero $\vec{\nabla}\times\vec{E}_p$ in (f). However, in comparison to (c), the curl is suppressed at all but the highest points. This is clear in the cross sections through (c) and (f) shown in (i). So the result of the procedure, (d)--(e), is closer to, but still short of, the true, curl-free $\vec{E}_p$.

Finally, we note that this method gives us no information about $\vec{E}_p$ outside the sample, where $J=0$. However, outside the strip the charge density $\rho/\epsilon_0 = \vec{\nabla}\cdot\vec{E_p} = 0$. Then $\vec{E_p}=-\vec{\nabla}\phi$ where $\phi$ obeys Laplace's equation, $\nabla^2 \phi=0$, with a Neumann boundary condition given by $\vec{E}_p$ in the strip.

\subsection{\label{ep_discussion}Discussion of $\vec{E}_p$ images}

Figure~\ref{p_fig}(c)--(j) shows the total electric field, $\vec{E} = \vec{E}_i + \vec{E}_p$, for the set of frames from Fig.~\ref{bjep}. The complete set of frames is assembled into the movie \texttt{EP.avi} in the supplemental material.\cite{mgroup_web} Our sample clearly deviates from a uniform strip, in which $\vec{E}_p=\vec{0}$. $\vec{E}_p$ is comparable in magnitude to $\vec{E}_i$, but much more inhomogeneous, contributing most at the spots identified by arrows in Fig.~\ref{bjep}(i) as weak points of the superconductor.

Thus we demonstrate reconstruction of the total electric field from our time-resolved magnetic images. We reiterate that this analysis is restricted to materials in which $\vec{E}$ is parallel to $\vec{J}$, and in which we can identify a field-free kernel.

\section{\label{power}Reconstructed power input}

\begin{figure*}[tb]
  \includegraphics[width=17.8cm]{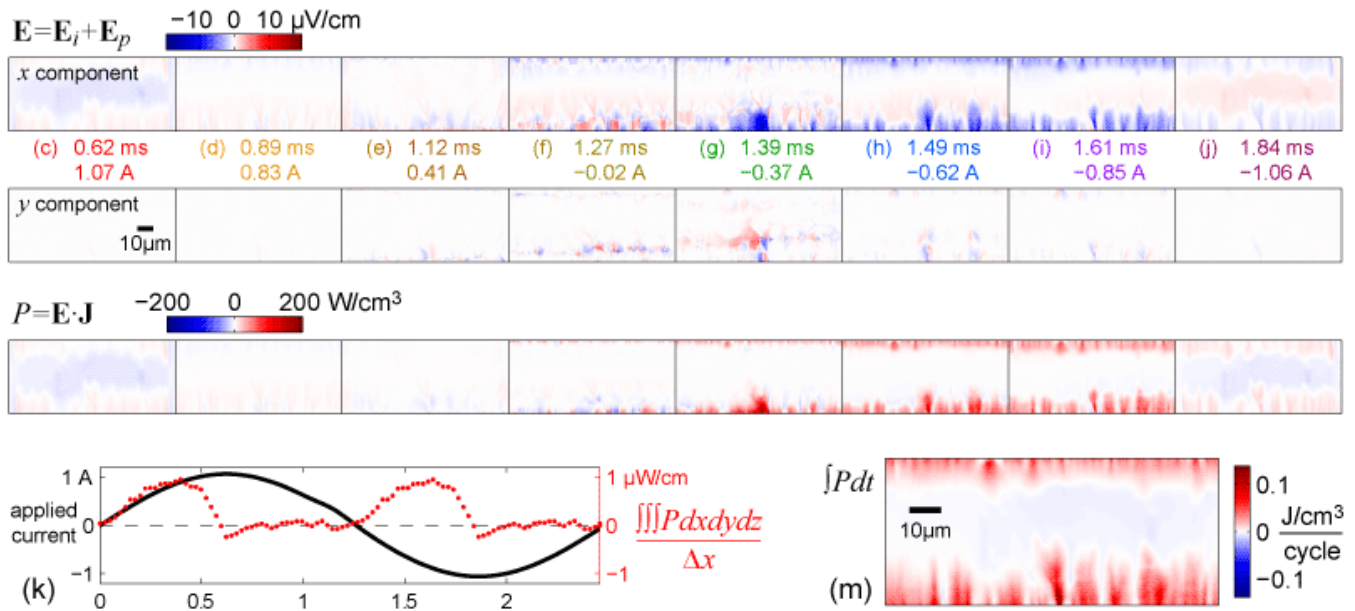}
  \caption{\label{p_fig}(Color online) (c)--(j) The total electric field, $\vec{E}$, and power, $P$, for the same set of frames shown in Fig.~\ref{bjep}. The images are cropped to the area of the strip. (k) Integrating $P$ over the area of each image and the thickness of the film, and normalizing by the length of strip imaged, we obtain the total power input as a function of time over a cycle of applied current. (m) Integrating $P$ over time, we see the spatial distribution of energy input over a cycle. Any reactive component of $P$ integrates to zero, leaving only the dissipated energy.}
\end{figure*}

Armed with $\vec{J}$ and $\vec{E}$, we calculate $P = \vec{J} \cdot \vec{E}$, the local power input to the film, resolved in time and space. The results are shown in the third row of Fig.~\ref{p_fig}. The largest features are the positive edge regions where vortices move in as the current sweeps. We note that the instantaneous power shown arises from both dissipation and reactance. These are not easily separated (e.g. by the relative phase of $J$ and $E$) because of the nonlinear relationship between $J$ and $E$.

We can relate our local measurements to macroscopic transport measurements by integrating over space, shown in Fig.~\ref{p_fig}(k). Each frame of the 50-frame set covering the central portion of the current cycle is summed and normalized by the length of strip imaged to obtain power per unit length, then plotted at its time within the cycle of applied current. These points are repeated in the first and fourth quarters of the cycle (in which $J$ and $E$ repeat with opposite signs). We note that when integrating over the entire sample (which we only do imperfectly by integrating over the image area), we expect no contribution to the power from the electrostatic field, $\vec{E}_p$, which, exerting a conservative force, cannot do work. Indeed, dropping $\vec{E}_p$ from the calculation shown in (k) does not change the result significantly.

The power input rises as the magnitude of current increases, but then falls back to become negative as the current reaches its peak. An inductive response with zero resistance would be zero at the peak. Instead, as discussed in Section~\ref{ei_discussion}, this negative contribution ($E$ opposite $J$) arises from relaxation of \jc. As the applied current decreases back to zero, we do not recover much power, as we would in a dissipation-free inductor, because the vortex movement is irreversible---the vortices remain pinned.

When integrating over time, any inductive contributions to the instantaneous power input cancel, leaving the sum of dissipation over one cycle, shown in Fig.~\ref{p_fig}(m). As in the instantaneous power, the edge contributions dominate. Integrating this image over space, or equivalently integrating the power in (k) over time, we obtain the energy dissipation per length of conductor, $6.6\cdot10^{-10}$~J/cm/cycle. For comparison, a calculation from the critical state model using the applied $I$ (1.07~A peak) and fit value of \ic (1.13~A) yields $1.2\cdot10^{-9}$~J/cm/cycle.\cite{brandt} Factors in the lower measured value may include suppression of high spatial frequencies in the reconstructed quantities due to regularization (see Section~\ref{j_method}) and cropping of the image, which may exclude some pixels near the edges of the strip.

This energy dissipation translates to an average power of $2.7\cdot10^{-7}$~W/cm, which, given the applied sample current of 0.75~Arms, implies a voltage of 6.9~nVrms between the voltage taps, which are spaced by 260~$\mu$m. We have not attempted to verify the presence of this small but perhaps measurable voltage.

\section{\label{ej}Local $E$--$J$ relations}

\begin{figure}[tb]
  \includegraphics[width=8.6cm]{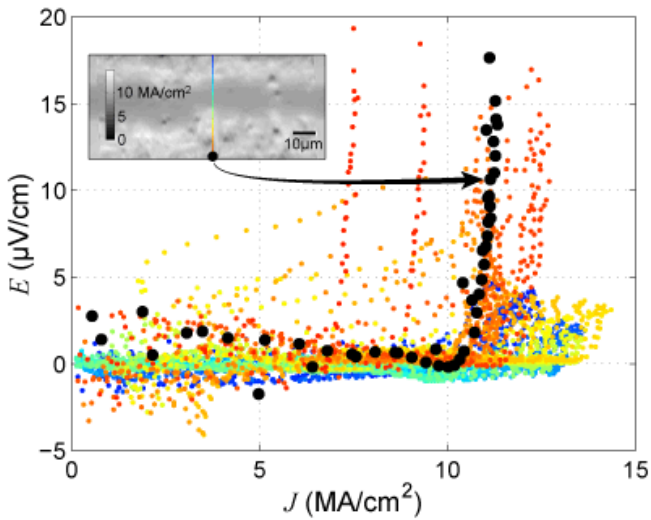}
  \caption{\label{fig:ej}(Color online) $E$, the magnitude of the projection of the total electric field onto $\vec{J}$, is plotted against the magnitude of $\vec{J}$. Values for one column of pixels are overlaid. The color of a set of points indicates its position on the bridge within the column of colored pixels in the inset, which is an image of the magnitude of $J$ at maximum applied current. The values for one pixel are highlighted in black and the pixel location is marked in the inset.}
\end{figure}

\begin{figure}[tb]
  \includegraphics[width=2.5in]{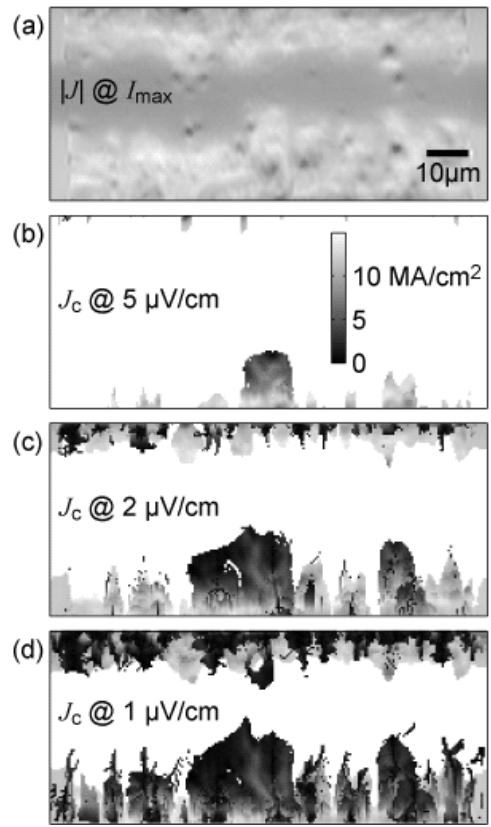}
  \caption{\label{Jc}(a) The magnitude of $J$ at maximum applied current compared to $J_c$ extracted from $E$--$J$ curves for various electric field criteria $E_c$, shown in (b)--(d). All are on the same color scale, shown in (b). Pixels that did not reach $E_c$ are white. Though (b)--(d) show more scatter, areas of low $J_c$ generally match those of low $J$ in (a).}
\end{figure}

Over the cycle of applied current, each location in the film experiences a range of current densities and electric field strengths. By plotting $E$ against $J$ for each pixel, we obtain a local characterization of the material. For example, the $E$--$J$ curve has often been used to test models of the flux pinning mechanism.\cite{gurevich_pinning,magnetic_relaxation_rmp,vortices_in_hts,beasley_creep,anderson-kim}

Figure~\ref{fig:ej} shows $E$--$J$ curves for one column of pixels at the locations indicated in the inset. One curve is highlighted, illustrating a plausible $E$--$J$ relation for a superconducting film. While the points at low $J$ display several $\mu$V/cm of scatter, there is a clear upturn at $J_c \approx 10$~MA/cm$^2$. This value is consistent with the 12~MA/cm$^2$ we observe in the edge regions of the strip at maximum applied current, and with 9~MA/cm$^2$ from transport measurements of a similar film at 44~K, as discussed in Section~\ref{j_discussion}.\cite{daniels_jc} However, as in Section~\ref{power}, we caution that the electric field we measure arises from both dissipation and reactance, and in comparing our data with a purely dissipative dc $E$--$J$ curve, we ignore reactance.

Although many of the curves seem reasonable, they deviate significantly between pixels, with upturns at current densities ranging from 14 to 2~MA/cm$^2$. These deviations are mapped out in Fig.~\ref{Jc} for various electric field criteria, $E_c$. The value of each pixel is the lowest $J$ for which $E_c$ is exceeded. If $E_c$ is not exceeded, the pixel is plotted as white. If we use Fig.~\ref{Jc}(a), which shows $J$ at maximum applied current, as an estimate of \Jc, it seems that many of the curves hit $E_c$ at erroneously low $J$. However, the spatial variation at least partially reflects genuine inhomogeneity of the material, because the points of low $J_c$ along the edges of (b)--(d) correlate with low points in (a).

\section{\label{err}Reconstruction errors}

\begin{figure*}[tb]
  \includegraphics[width=17.8cm]{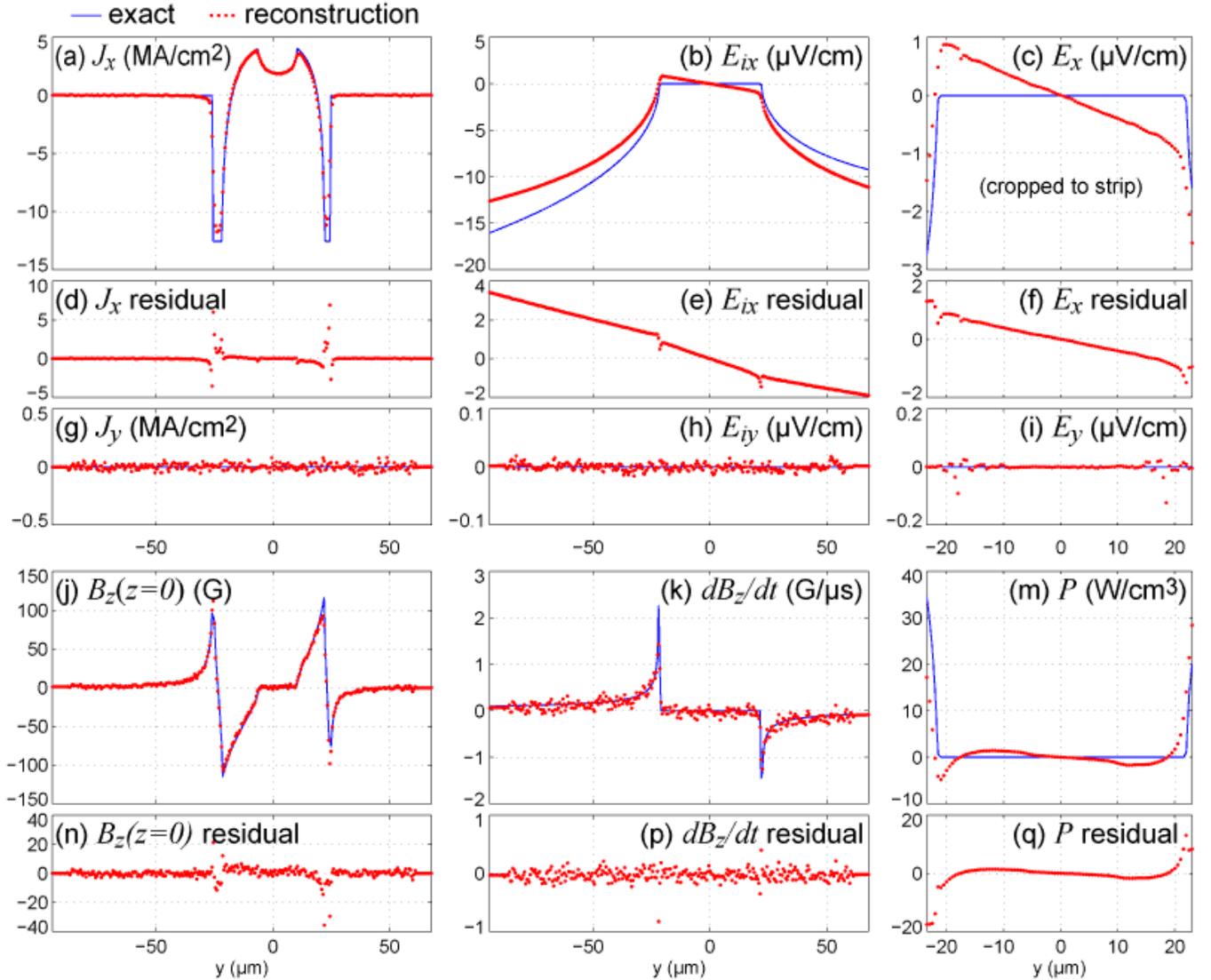}
  \caption{\label{err_fig}(Color online) Errors introduced by the reconstruction procedure are analyzed by executing the procedure on model data. The exact quantities, known from the model, are plotted as blue lines. From them, the magnetic field in the measurement plane and its time derivative are calculated, noise is added, and the quantities are reconstructed. The red dots are $y$ cross sections through the centers of the reconstructed images. The residuals are the reconstructed values minus the exact values. The only variation as a function of $x$ comes from the noise, so the exact values of the $y$ components are zero.}
\end{figure*}

Throughout the paper we have tried to point out physical assumptions and where they may break down. Even if our assumptions hold, however, the reconstruction procedure can introduce errors and amplify uncertainties. The magnitudes of these effects are difficult to predict analytically because of the many complex numerical transformations involved. Instead, we estimate the errors by executing the procedure on simulated data, for which we know exact solutions to compare to the reconstructed quantities. Figure~\ref{err_fig} compares cross sections from the resultant exact and reconstructed images.

We generate the data with the same critical state model used for the background of the magnetic images, as described in Section~\ref{b}.\cite{brandt,strip_model} The parameters are the same, with the applied current (-0.02~A) corresponding to Fig.~\ref{bjep}(f). We choose this frame because it includes regions of zero current, which present a worst case for reconstructing $\vec{E}_p$, as described below.

As the input to the reconstruction procedure, we calculate an image of the magnetic field in the measurement plane (z=1.2~$\mu$m). This is identical to the background of Fig.~\ref{bjep}(f), but extends over the entire image area. As in the background, we account for the sensor's tilt (3\dg about the $x$ axis), which adds a small portion of $-B_y$ to the measured ``$B_z$.'' We then add normally distributed noise within the measurement region (the subset of the image defined by the dashed green box in Fig.~\ref{bjep}(c)). The amplitude of the noise matches that observed in the real data outside the strip.

Using the reconstruction procedure, we obtain $J_x$, $J_y$, and $B_z(z=0)$, shown in the first column of Fig.~\ref{err_fig}. The reconstructed $J_x$ smooths the sharp corners of the true $J_x$. It also displays 0.03~MA/cm$^2$~rms of noise, as does the reconstructed $J_y$. The only variation as a function of $x$ comes from the noise added to $B$; the exact values of all $y$ components (and $E_{px}$) are zero.

To reconstruct $\vec{E}$, we similarly calculate $B$ for applied currents corresponding to 12.5~$\mu$s before and after Fig.~\ref{bjep}(f). From each $B$ image we reconstruct $B_z(z=0)$, and subtract to approximate $dB_z(z=0)/dt$, shown in Fig.~\ref{err_fig}(k). We then follow the reconstruction procedure for $\vec{E}_i$ (as in Section~\ref{ei}) and $\vec{E}_p$ (Section~\ref{ep}).

We see in Fig.~\ref{err_fig}(e) that an erroneous negative slope is present in the residual of the reconstructed $E_{ix}$. This error can be traced to the small portion of $-B_y$ in $B$, which is not accounted for by the reconstruction procedure. In principle, the procedure could be adapted to assume a specified linear combination of $B_y$ and $B_z$, though relations such as Eq.~\ref{gb_eq} would become more complicated. One could also calculate an approximate $B_y$ from the reconstructed $J_x$, subtract it from the measured $B$, then iterate, reconstructing a more accurate $J_x$ and $B_z$ each iteration. This procedure has been successfully applied to removing in-plane field components from magneto-optic images.\cite{mo_bx} Both procedures require precise knowledge of the sensor tilt.

Although small compared to the error in $E_{ix}$, other errors are worth noting. For one, we compare the $y$ components of the inductive electric field and total electric field (Fig.~\ref{err_fig}(h) versus (i)). The addition of $E_{py}$ suppresses $E_y$ (from about $6\cdot10^{-3}$ to $2\cdot10^{-3}$~$\mu$V/cm) everywhere except near $y=\pm18$~$\mu$m, where $E_y$ fluctuates with $x$ by 0.1~$\mu$V/cm rms over the image. At these locations, $J$ is approximately zero, making the direction of $\vec{J}$ completely uncertain, which renders Eqs.~\ref{eiep}--\ref{integral_ep} inaccurate. Furthermore, such errors may scale up with the larger $E_p$ present in the real data. Thus we must admit $\sim100\%$ uncertainty in $E_p$ near regions in which $J$ drops below its noise level, 0.03~MA/cm$^2$. More generally, we point out that this error analysis is performed for a homogeneous model, which may not account for some features of real, inhomogeneous systems.

\subsection{\label{curlej}Alternate reconstruction of $E$ from $E_i$}

Where there are large uncertainties in $\vec{E}_p$, one can use an alternative route to reconstructing $\vec{E}$ that bypasses $\vec{E}_p$. This method starts with $\vec{\nabla}\times\vec{E}$, to which $\vec{E}_p$ (which is curl free by definition) does not contribute. As in Section~\ref{ep}, we assume $\vec{E}$ is parallel to $\vec{J}$, in which case we can write
\begin{equation}
\vec{E} = \rho(x,y,J) \vec{J}
\end{equation}
and describe $\vec{E}$ via the scalar $\rho$ (the local resistivity). Swapping sides and taking the curl, we have
\begin{equation}
(\vec{\nabla}\rho)\times\vec{J} + \rho (\vec{\nabla}\times\vec{J}) = \vec{\nabla}\times\vec{E}
\end{equation}
Taking the $z$ component and applying Faraday's law (Eq.~\ref{ecurl}) to the right-hand side,
\begin{equation}
(\partial_x \rho) J_y - (\partial_y \rho) J_x + \rho (\partial_x J_y - \partial_y Jx) = -\partial_t B_z
\end{equation}
To proceed, we assume that the spatial dependence of $\rho$ arises solely from its dependence on $J$, i.e. that the $E$--$J$ relation is constant over the single-pixel scale at which we solve this equation. Then,
\begin{equation}
\frac{d\rho}{dJ} \left[(\partial_x J)J_y - (\partial_y J)J_x\right] + \rho (\partial_x J_y - \partial_y Jx) = -\partial_t B_z
\end{equation}
which we can write in a standard form
\begin{equation}
\frac{d\rho}{dJ} + \rho p(J) = q(J)
\end{equation}
where
\begin{equation}
p(J) \equiv \frac{\partial_x J_y - \partial_y Jx}{(\partial_x J)J_y - (\partial_y J)J_x}
\mbox{ and } q(J) \equiv \frac{-\partial_t B_z}{(\partial_x J)J_y - (\partial_y J)J_x}
\end{equation}
are known. The solution to this differential equation is\cite{dedj}
\begin{equation}
\rho(J) = \frac{\int^{J}_{0}u(J)q(J)dJ}{u(J)}
\label{solution}
\end{equation}
where $u \equiv \mathrm{exp}\left(\int^{J}_{0}p(J)dJ\right)$. We use the boundary condition $\rho(0)=0$.

So the ingredients are the reconstructed $\partial_t B_z(z=0)$ and $\vec{J}$, along with its spatial derivatives. These quantities are arranged by time; we re-sort them by $J$. Then the integrals with respect to $J$ are calculated by the trapezoid method. Applying Eq.~\ref{solution} at each pixel, we obtain $\rho(J)$ and therefore $\vec{E}(J)$ at each point. We could then reconstruct $\vec{E}_p=\vec{E}-\vec{E}_i$ if desired.

In practice, our results are dominated by noise. In our images, the spatial derivatives of $J$ are of order $10^{-3}$~MA/cm$^3$ while their uncertainties (based on the 0.03~MA/cm$^2$ uncertainty in $J$) are several orders of magnitude larger. This method may be effective, however, for other samples or measurement techniques.

\section{Conclusion}
We have shown that time-resolved magnetic imaging of a superconducting thin film yields a complete characterization of its electromagnetic properties, including distributions of current flow, electric field, power, and local $E$--$J$ relations. We also point out the physical assumptions and requirements behind the mathematical transformations. The technique is compatible with various methods of magnetic imaging. It requires time resolution commensurate with the sample conditions being studied, but this capability is not limited to scanning Hall probe microscopes. Magneto-optics can use high-speed frame grabbing or a phase-locked short pulse technique to acquire similar data sets (albeit in a different pixel sequence).\cite{barnes} Alternatively, the average response can be acquired from each pixel in succession, as is done here, at the expense of longer total acquisition time.

An applied field, $B_z(t)$, can be substituted for our applied current, and its time dependence need not be sinusoidal. For example, the analysis is applicable to the transient response following a change in applied current or field, as in Ref.~\onlinecite{jooss}. However, by demonstrating the technique on a superconducting strip carrying an applied current at 400~Hz, a realistic operating frequency for power applications,\cite{400Hz} we point out that an important use of this work is the characterization of material inhomogeneity and its effect on ac losses. Such imaging may also provide an efficient method for measuring the effects of complex sample geometries.\cite{roebel}

As mentioned in Section~\ref{ej}, $E$--$J$ curves give insight into the flux pinning mechanism.\cite{gurevich_pinning,magnetic_relaxation_rmp,vortices_in_hts,beasley_creep,anderson-kim} By resolving the variation in $E$ versus $J$ with position and time, this analysis may allow one to dissect the behavior of a heterogeneous sample, correlating pinning dynamics with materials properties. One may also examine correlations with quantities, such as magnetic field strength, that vary with space or time, fully accounting for the heterogeneous self-field experienced by different parts of the sample.

\begin{acknowledgments}
We gratefully acknowledge several contributions: The Hall sensor was fabricated by Janice Wynn Guikema using GaAs/AlGaAs heterostructures grown by David Kisker. Joseph Sulpizio and Hung-Tao Chou grew the sensor's oxide layer. The YBCO film was grown by George Daniels. Christian Jooss lent useful advice and experience. This work was supported by the Air Force Office of Scientific Research through a Multi-University Research Initiative (MURI).
\end{acknowledgments}


\newpage
\begin{thebibliography}{38}
\expandafter\ifx\csname natexlab\endcsname\relax\def\natexlab#1{#1}\fi
\expandafter\ifx\csname bibnamefont\endcsname\relax
  \def\bibnamefont#1{#1}\fi
\expandafter\ifx\csname bibfnamefont\endcsname\relax
  \def\bibfnamefont#1{#1}\fi
\expandafter\ifx\csname citenamefont\endcsname\relax
  \def\citenamefont#1{#1}\fi
\expandafter\ifx\csname url\endcsname\relax
  \def\url#1{\texttt{#1}}\fi
\expandafter\ifx\csname urlprefix\endcsname\relax\def\urlprefix{URL }\fi
\providecommand{\bibinfo}[2]{#2}
\providecommand{\eprint}[2][]{\url{#2}}

\bibitem[{\citenamefont{Larbalestier et~al.}(2001)\citenamefont{Larbalestier,
  Gurevich, Feldmann, and Polyanskii}}]{cc}
\bibinfo{author}{\bibfnamefont{D.}~\bibnamefont{Larbalestier}},
  \bibinfo{author}{\bibfnamefont{A.}~\bibnamefont{Gurevich}},
  \bibinfo{author}{\bibfnamefont{D.~M.} \bibnamefont{Feldmann}},
  \bibnamefont{and}
  \bibinfo{author}{\bibfnamefont{A.}~\bibnamefont{Polyanskii}},
  \bibinfo{journal}{Nature} \textbf{\bibinfo{volume}{414}},
  \bibinfo{pages}{368} (\bibinfo{year}{2001}).

\bibitem[{\citenamefont{Roth et~al.}(1989)\citenamefont{Roth, Sepulveda, and
  Wikswo}}]{wikswo}
\bibinfo{author}{\bibfnamefont{B.~J.} \bibnamefont{Roth}},
  \bibinfo{author}{\bibfnamefont{N.~G.} \bibnamefont{Sepulveda}},
  \bibnamefont{and} \bibinfo{author}{\bibfnamefont{J.~P.}
  \bibnamefont{Wikswo}}, \bibinfo{journal}{J. Appl. Phys.}
  \textbf{\bibinfo{volume}{65}}, \bibinfo{pages}{361 } (\bibinfo{year}{1989}).

\bibitem[{\citenamefont{Wijngaarden et~al.}(1998)\citenamefont{Wijngaarden,
  Heeck, Spoelder, Surdeanu, and Griessen}}]{cg}
\bibinfo{author}{\bibfnamefont{R.~J.} \bibnamefont{Wijngaarden}},
  \bibinfo{author}{\bibfnamefont{K.}~\bibnamefont{Heeck}},
  \bibinfo{author}{\bibfnamefont{H.~J.~W.} \bibnamefont{Spoelder}},
  \bibinfo{author}{\bibfnamefont{R.}~\bibnamefont{Surdeanu}}, \bibnamefont{and}
  \bibinfo{author}{\bibfnamefont{R.}~\bibnamefont{Griessen}},
  \bibinfo{journal}{Physica C} \textbf{\bibinfo{volume}{295}},
  \bibinfo{pages}{177 } (\bibinfo{year}{1998}).

\bibitem[{\citenamefont{Feldmann}(2004)}]{reg}
\bibinfo{author}{\bibfnamefont{D.~M.} \bibnamefont{Feldmann}},
  \bibinfo{journal}{Phys. Rev. B} \textbf{\bibinfo{volume}{69}},
  \bibinfo{pages}{144515} (\bibinfo{year}{2004}).

\bibitem[{\citenamefont{Patnaik et~al.}(2003)\citenamefont{Patnaik, Feldmann,
  Polyanskii, Yuan, Jiang, Cai, Hellstrom, Larbalestier, and
  Huang}}]{feldmann_mo_cr}
\bibinfo{author}{\bibfnamefont{S.}~\bibnamefont{Patnaik}},
  \bibinfo{author}{\bibfnamefont{D.~M.} \bibnamefont{Feldmann}},
  \bibinfo{author}{\bibfnamefont{A.~A.} \bibnamefont{Polyanskii}},
  \bibinfo{author}{\bibfnamefont{Y.}~\bibnamefont{Yuan}},
  \bibinfo{author}{\bibfnamefont{J.}~\bibnamefont{Jiang}},
  \bibinfo{author}{\bibfnamefont{X.~Y.} \bibnamefont{Cai}},
  \bibinfo{author}{\bibfnamefont{E.~E.} \bibnamefont{Hellstrom}},
  \bibinfo{author}{\bibfnamefont{D.~C.} \bibnamefont{Larbalestier}},
  \bibnamefont{and} \bibinfo{author}{\bibfnamefont{Y.}~\bibnamefont{Huang}},
  \bibinfo{journal}{IEEE T. Appl. Supercon.} \textbf{\bibinfo{volume}{13}},
  \bibinfo{pages}{2930 } (\bibinfo{year}{2003}).

\bibitem[{\citenamefont{Jooss et~al.}(2002)\citenamefont{Jooss, Albrecht, Kuhn,
  Leonhardt, and Kronmuller}}]{jooss_review}
\bibinfo{author}{\bibfnamefont{C.}~\bibnamefont{Jooss}},
  \bibinfo{author}{\bibfnamefont{J.}~\bibnamefont{Albrecht}},
  \bibinfo{author}{\bibfnamefont{H.}~\bibnamefont{Kuhn}},
  \bibinfo{author}{\bibfnamefont{S.}~\bibnamefont{Leonhardt}},
  \bibnamefont{and}
  \bibinfo{author}{\bibfnamefont{H.}~\bibnamefont{Kronmuller}},
  \bibinfo{journal}{Rep. Prog. Phys.} \textbf{\bibinfo{volume}{65}},
  \bibinfo{pages}{651 } (\bibinfo{year}{2002}).

\bibitem[{\citenamefont{Koblischka and
  Wijngaarden}(1995)}]{wijngaarden_mo_review}
\bibinfo{author}{\bibfnamefont{M.~R.} \bibnamefont{Koblischka}}
  \bibnamefont{and} \bibinfo{author}{\bibfnamefont{R.~J.}
  \bibnamefont{Wijngaarden}}, \bibinfo{journal}{Supercon. Sci. Technol.}
  \textbf{\bibinfo{volume}{8}}, \bibinfo{pages}{199 } (\bibinfo{year}{1995}).

\bibitem[{\citenamefont{Dinner et~al.}(2005)\citenamefont{Dinner, Beasley, and
  Moler}}]{rsi}
\bibinfo{author}{\bibfnamefont{R.~B.} \bibnamefont{Dinner}},
  \bibinfo{author}{\bibfnamefont{M.~R.} \bibnamefont{Beasley}},
  \bibnamefont{and} \bibinfo{author}{\bibfnamefont{K.~A.} \bibnamefont{Moler}},
  \bibinfo{journal}{Rev. Sci. Instrum.} \textbf{\bibinfo{volume}{76}},
  \bibinfo{pages}{103702 } (\bibinfo{year}{2005}).

\bibitem[{\citenamefont{Grant et~al.}(1994)\citenamefont{Grant, Denhoff, Xing,
  Brown, Govorkov, Irwin, Heinrich, Zhou, Fife, and Cragg}}]{shp_mo_cr}
\bibinfo{author}{\bibfnamefont{P.~D.} \bibnamefont{Grant}},
  \bibinfo{author}{\bibfnamefont{M.~W.} \bibnamefont{Denhoff}},
  \bibinfo{author}{\bibfnamefont{W.}~\bibnamefont{Xing}},
  \bibinfo{author}{\bibfnamefont{P.}~\bibnamefont{Brown}},
  \bibinfo{author}{\bibfnamefont{S.}~\bibnamefont{Govorkov}},
  \bibinfo{author}{\bibfnamefont{J.~C.} \bibnamefont{Irwin}},
  \bibinfo{author}{\bibfnamefont{B.}~\bibnamefont{Heinrich}},
  \bibinfo{author}{\bibfnamefont{H.}~\bibnamefont{Zhou}},
  \bibinfo{author}{\bibfnamefont{A.~A.} \bibnamefont{Fife}}, \bibnamefont{and}
  \bibinfo{author}{\bibfnamefont{A.~R.} \bibnamefont{Cragg}},
  \bibinfo{journal}{Physica C} \textbf{\bibinfo{volume}{229}},
  \bibinfo{pages}{289 } (\bibinfo{year}{1994}).

\bibitem[{\citenamefont{Inoue et~al.}(2005)\citenamefont{Inoue, Kiss, Koyanagi,
  Imamura, Takeo, Iijima, Kakimoto, Saitoh, Matsuda, Tokunaga
  et~al.}}]{kiss_squid_slm}
\bibinfo{author}{\bibfnamefont{M.}~\bibnamefont{Inoue}},
  \bibinfo{author}{\bibfnamefont{T.}~\bibnamefont{Kiss}},
  \bibinfo{author}{\bibfnamefont{S.}~\bibnamefont{Koyanagi}},
  \bibinfo{author}{\bibfnamefont{K.}~\bibnamefont{Imamura}},
  \bibinfo{author}{\bibfnamefont{M.}~\bibnamefont{Takeo}},
  \bibinfo{author}{\bibfnamefont{Y.}~\bibnamefont{Iijima}},
  \bibinfo{author}{\bibfnamefont{K.}~\bibnamefont{Kakimoto}},
  \bibinfo{author}{\bibfnamefont{T.}~\bibnamefont{Saitoh}},
  \bibinfo{author}{\bibfnamefont{J.}~\bibnamefont{Matsuda}},
  \bibinfo{author}{\bibfnamefont{Y.}~\bibnamefont{Tokunaga}},
  \bibnamefont{et~al.}, \bibinfo{journal}{Physica C}
  \textbf{\bibinfo{volume}{426}}, \bibinfo{pages}{1068 }
  (\bibinfo{year}{2005}).

\bibitem[{\citenamefont{Kirtley et~al.}(1995)\citenamefont{Kirtley, Ketchen,
  Stawiasz, Sun, Gallagher, Blanton, and Wind}}]{ssquid}
\bibinfo{author}{\bibfnamefont{J.~R.} \bibnamefont{Kirtley}},
  \bibinfo{author}{\bibfnamefont{M.~B.} \bibnamefont{Ketchen}},
  \bibinfo{author}{\bibfnamefont{K.~G.} \bibnamefont{Stawiasz}},
  \bibinfo{author}{\bibfnamefont{J.~Z.} \bibnamefont{Sun}},
  \bibinfo{author}{\bibfnamefont{W.~J.} \bibnamefont{Gallagher}},
  \bibinfo{author}{\bibfnamefont{S.~H.} \bibnamefont{Blanton}},
  \bibnamefont{and} \bibinfo{author}{\bibfnamefont{S.~J.} \bibnamefont{Wind}},
  \bibinfo{journal}{Appl. Phys. Lett.} \textbf{\bibinfo{volume}{66}},
  \bibinfo{pages}{1138} (\bibinfo{year}{1995}).

\bibitem[{\citenamefont{Perkins et~al.}(2001)\citenamefont{Perkins,
  Bugoslavsky, and Caplin}}]{potentiometry_on_cc}
\bibinfo{author}{\bibfnamefont{G.~K.} \bibnamefont{Perkins}},
  \bibinfo{author}{\bibfnamefont{Y.~V.} \bibnamefont{Bugoslavsky}},
  \bibnamefont{and} \bibinfo{author}{\bibfnamefont{A.~D.}
  \bibnamefont{Caplin}}, \bibinfo{journal}{Supercon. Sci. Technol.}
  \textbf{\bibinfo{volume}{14}}, \bibinfo{pages}{685 } (\bibinfo{year}{2001}).

\bibitem[{\citenamefont{Abraimov et~al.}(2004)\citenamefont{Abraimov, Feldmann,
  Polyanskii, Gurevich, Daniels, Larbalestier, Zhuravel, and Ustinov}}]{ltslm}
\bibinfo{author}{\bibfnamefont{D.}~\bibnamefont{Abraimov}},
  \bibinfo{author}{\bibfnamefont{D.~M.} \bibnamefont{Feldmann}},
  \bibinfo{author}{\bibfnamefont{A.~A.} \bibnamefont{Polyanskii}},
  \bibinfo{author}{\bibfnamefont{A.}~\bibnamefont{Gurevich}},
  \bibinfo{author}{\bibfnamefont{G.}~\bibnamefont{Daniels}},
  \bibinfo{author}{\bibfnamefont{D.~C.} \bibnamefont{Larbalestier}},
  \bibinfo{author}{\bibfnamefont{A.~P.} \bibnamefont{Zhuravel}},
  \bibnamefont{and} \bibinfo{author}{\bibfnamefont{A.~V.}
  \bibnamefont{Ustinov}}, \bibinfo{journal}{Applied Physics Letters}
  \textbf{\bibinfo{volume}{85}}, \bibinfo{pages}{2568 } (\bibinfo{year}{2004}).

\bibitem[{\citenamefont{Gross and Koelle}(1994)}]{ltsem}
\bibinfo{author}{\bibfnamefont{R.}~\bibnamefont{Gross}} \bibnamefont{and}
  \bibinfo{author}{\bibfnamefont{D.}~\bibnamefont{Koelle}},
  \bibinfo{journal}{Rep. Prog. Phys.} \textbf{\bibinfo{volume}{57}},
  \bibinfo{pages}{651 } (\bibinfo{year}{1994}).

\bibitem[{\citenamefont{Norris}(1970)}]{norris}
\bibinfo{author}{\bibfnamefont{W.~T.} \bibnamefont{Norris}},
  \bibinfo{journal}{J. Phys. D} \textbf{\bibinfo{volume}{3}},
  \bibinfo{pages}{489 } (\bibinfo{year}{1970}).

\bibitem[{\citenamefont{Brandt and Indenbom}(1993)}]{brandt}
\bibinfo{author}{\bibfnamefont{E.~H.} \bibnamefont{Brandt}} \bibnamefont{and}
  \bibinfo{author}{\bibfnamefont{M.}~\bibnamefont{Indenbom}},
  \bibinfo{journal}{Phys. Rev. B} \textbf{\bibinfo{volume}{48}},
  \bibinfo{pages}{12893 } (\bibinfo{year}{1993}).

\bibitem[{\citenamefont{Jooss and Born}(2006)}]{jooss}
\bibinfo{author}{\bibfnamefont{C.}~\bibnamefont{Jooss}} \bibnamefont{and}
  \bibinfo{author}{\bibfnamefont{V.}~\bibnamefont{Born}},
  \bibinfo{journal}{Phys. Rev. B} \textbf{\bibinfo{volume}{73}},
  \bibinfo{pages}{94508 } (\bibinfo{year}{2006}).

\bibitem[{\citenamefont{Kalsi}(2004)}]{400Hz}
\bibinfo{author}{\bibfnamefont{S.}~\bibnamefont{Kalsi}},
  \bibinfo{journal}{Proc. IEEE} \textbf{\bibinfo{volume}{92}},
  \bibinfo{pages}{1688} (\bibinfo{year}{2004}).

\bibitem[{ald()}]{ald}
\bibinfo{note}{The insulating layer consists of 50~nm of Al$_2$O$_3$ grown via
  atomic layer deposition using a Cambridge NanoTech Inc. Savannah 200. The
  growth occurred by depositing trimethylaluminum, oxidizing it with water, and
  iterating 500 times.}

\bibitem[{epa()}]{mgroup_web}
\bibinfo{note}{Movies and other supplementary materials are available at \url{http://www.stanford.edu/group/moler/rdinner}}

\bibitem[{\citenamefont{Zeldov et~al.}(1994)\citenamefont{Zeldov, Clem,
  McElfresh, and Darwin}}]{strip_model}
\bibinfo{author}{\bibfnamefont{E.}~\bibnamefont{Zeldov}},
  \bibinfo{author}{\bibfnamefont{J.~R.} \bibnamefont{Clem}},
  \bibinfo{author}{\bibfnamefont{M.}~\bibnamefont{McElfresh}},
  \bibnamefont{and} \bibinfo{author}{\bibfnamefont{M.}~\bibnamefont{Darwin}},
  \bibinfo{journal}{Phys. Rev. B} \textbf{\bibinfo{volume}{49}},
  \bibinfo{pages}{9802} (\bibinfo{year}{1994}).

\bibitem[{\citenamefont{Djordjevic et~al.}(2002)\citenamefont{Djordjevic,
  Farber, Deutscher, Bontemps, Durand, and Contour}}]{ybco_penetration_depth}
\bibinfo{author}{\bibfnamefont{S.}~\bibnamefont{Djordjevic}},
  \bibinfo{author}{\bibfnamefont{E.}~\bibnamefont{Farber}},
  \bibinfo{author}{\bibfnamefont{G.}~\bibnamefont{Deutscher}},
  \bibinfo{author}{\bibfnamefont{N.}~\bibnamefont{Bontemps}},
  \bibinfo{author}{\bibfnamefont{O.}~\bibnamefont{Durand}}, \bibnamefont{and}
  \bibinfo{author}{\bibfnamefont{J.}~\bibnamefont{Contour}},
  \bibinfo{journal}{Eur. Phys. J. B} \textbf{\bibinfo{volume}{25}},
  \bibinfo{pages}{407} (\bibinfo{year}{2002}).

\bibitem[{\citenamefont{Carneiro and Brandt}(2000)}]{vortex_thick_film}
\bibinfo{author}{\bibfnamefont{G.}~\bibnamefont{Carneiro}} \bibnamefont{and}
  \bibinfo{author}{\bibfnamefont{E.~H.} \bibnamefont{Brandt}},
  \bibinfo{journal}{Phys. Rev. B} \textbf{\bibinfo{volume}{61}},
  \bibinfo{pages}{6370 } (\bibinfo{year}{2000}).

\bibitem[{\citenamefont{Brandt}(1995)}]{brandt_prl}
\bibinfo{author}{\bibfnamefont{E.~H.} \bibnamefont{Brandt}},
  \bibinfo{journal}{Phys. Rev. Lett.} \textbf{\bibinfo{volume}{74}},
  \bibinfo{pages}{3025 } (\bibinfo{year}{1995}).

\bibitem[{\citenamefont{Bending}(1999)}]{bending_review}
\bibinfo{author}{\bibfnamefont{S.~J.} \bibnamefont{Bending}},
  \bibinfo{journal}{Adv. Phys.} \textbf{\bibinfo{volume}{48}},
  \bibinfo{pages}{449} (\bibinfo{year}{1999}).

\bibitem[{\citenamefont{Daniels et~al.}(2000)\citenamefont{Daniels, Gurevich,
  and Larbalestier}}]{daniels_jc}
\bibinfo{author}{\bibfnamefont{G.~A.} \bibnamefont{Daniels}},
  \bibinfo{author}{\bibfnamefont{A.}~\bibnamefont{Gurevich}}, \bibnamefont{and}
  \bibinfo{author}{\bibfnamefont{D.~C.} \bibnamefont{Larbalestier}},
  \bibinfo{journal}{Appl. Phys. Lett.} \textbf{\bibinfo{volume}{77}},
  \bibinfo{pages}{3251 } (\bibinfo{year}{2000}).

\bibitem[{joo()}]{jooss_incorrect}
\bibinfo{note}{Reference~\onlinecite{jooss} makes a physically motivated
  argument that $E_{p\|}$ is approximately zero for the case of magnetization
  decay. This argument is incorrect and leads to a significant curl in
  $\vec{E}_p$, visible in Ref.~\onlinecite{jooss}'s Fig.~12 as $-\partial_y
  E_{px} + \partial_x E_{py} \neq 0$.}

\bibitem[{\citenamefont{Kunchur et~al.}(1994)\citenamefont{Kunchur, Christen,
  Klabunde, and Phillips}}]{ybco_hall}
\bibinfo{author}{\bibfnamefont{M.~N.} \bibnamefont{Kunchur}},
  \bibinfo{author}{\bibfnamefont{D.~K.} \bibnamefont{Christen}},
  \bibinfo{author}{\bibfnamefont{C.~E.} \bibnamefont{Klabunde}},
  \bibnamefont{and} \bibinfo{author}{\bibfnamefont{J.~M.}
  \bibnamefont{Phillips}}, \bibinfo{journal}{Phys. Rev. Lett.}
  \textbf{\bibinfo{volume}{72}}, \bibinfo{pages}{2259 } (\bibinfo{year}{1994}).

\bibitem[{\citenamefont{Gurevich}(2006)}]{gurevich_talk}
\bibinfo{author}{\bibfnamefont{A.}~\bibnamefont{Gurevich}}
  (\bibinfo{year}{2006}), \bibinfo{note}{talk at the Stanford-Wisconsin Coated
  Conductor Workshop}.

\bibitem[{\citenamefont{Gurevich}(1990)}]{gurevich_pinning}
\bibinfo{author}{\bibfnamefont{A.}~\bibnamefont{Gurevich}},
  \bibinfo{journal}{Phys. Rev. B} \textbf{\bibinfo{volume}{42}},
  \bibinfo{pages}{4857} (\bibinfo{year}{1990}).

\bibitem[{\citenamefont{Yeshurun et~al.}(1996)\citenamefont{Yeshurun,
  Malozemoff, and Shaulov}}]{magnetic_relaxation_rmp}
\bibinfo{author}{\bibfnamefont{Y.}~\bibnamefont{Yeshurun}},
  \bibinfo{author}{\bibfnamefont{A.~P.} \bibnamefont{Malozemoff}},
  \bibnamefont{and} \bibinfo{author}{\bibfnamefont{A.}~\bibnamefont{Shaulov}},
  \bibinfo{journal}{Rev. Mod. Phys.} \textbf{\bibinfo{volume}{68}},
  \bibinfo{pages}{911} (\bibinfo{year}{1996}).

\bibitem[{\citenamefont{Blatter et~al.}(1994)\citenamefont{Blatter, Feigel'man,
  Geshkenbein, Larkin, and Vinokur}}]{vortices_in_hts}
\bibinfo{author}{\bibfnamefont{G.}~\bibnamefont{Blatter}},
  \bibinfo{author}{\bibfnamefont{M.~V.} \bibnamefont{Feigel'man}},
  \bibinfo{author}{\bibfnamefont{V.~B.} \bibnamefont{Geshkenbein}},
  \bibinfo{author}{\bibfnamefont{A.~I.} \bibnamefont{Larkin}},
  \bibnamefont{and} \bibinfo{author}{\bibfnamefont{V.~M.}
  \bibnamefont{Vinokur}}, \bibinfo{journal}{Rev. Mod. Phys.}
  \textbf{\bibinfo{volume}{66}}, \bibinfo{pages}{1125} (\bibinfo{year}{1994}).

\bibitem[{\citenamefont{Beasley et~al.}(1969)\citenamefont{Beasley, Labusch,
  and Webb}}]{beasley_creep}
\bibinfo{author}{\bibfnamefont{M.~R.} \bibnamefont{Beasley}},
  \bibinfo{author}{\bibfnamefont{R.}~\bibnamefont{Labusch}}, \bibnamefont{and}
  \bibinfo{author}{\bibfnamefont{W.~W.} \bibnamefont{Webb}},
  \bibinfo{journal}{Phys. Rev.} \textbf{\bibinfo{volume}{181}},
  \bibinfo{pages}{682} (\bibinfo{year}{1969}).

\bibitem[{\citenamefont{Anderson and Kim}(1964)}]{anderson-kim}
\bibinfo{author}{\bibfnamefont{P.~W.} \bibnamefont{Anderson}} \bibnamefont{and}
  \bibinfo{author}{\bibfnamefont{Y.~B.} \bibnamefont{Kim}},
  \bibinfo{journal}{Rev. Mod. Phys.} \textbf{\bibinfo{volume}{36}},
  \bibinfo{pages}{39} (\bibinfo{year}{1964}).

\bibitem[{\citenamefont{Laviano et~al.}(2003)\citenamefont{Laviano, Botta,
  Chiodoni, Gerbaldo, Ghigo, Gozzelino, Zannella, and Mezzetti}}]{mo_bx}
\bibinfo{author}{\bibfnamefont{F.}~\bibnamefont{Laviano}},
  \bibinfo{author}{\bibfnamefont{D.}~\bibnamefont{Botta}},
  \bibinfo{author}{\bibfnamefont{A.}~\bibnamefont{Chiodoni}},
  \bibinfo{author}{\bibfnamefont{R.}~\bibnamefont{Gerbaldo}},
  \bibinfo{author}{\bibfnamefont{G.}~\bibnamefont{Ghigo}},
  \bibinfo{author}{\bibfnamefont{L.}~\bibnamefont{Gozzelino}},
  \bibinfo{author}{\bibfnamefont{S.}~\bibnamefont{Zannella}}, \bibnamefont{and}
  \bibinfo{author}{\bibfnamefont{E.}~\bibnamefont{Mezzetti}},
  \bibinfo{journal}{Supercon. Sci. Technol.} \textbf{\bibinfo{volume}{16}},
  \bibinfo{pages}{71 } (\bibinfo{year}{2003}).

\bibitem[{\citenamefont{Rainville and Bedient}(1969)}]{dedj}
\bibinfo{author}{\bibfnamefont{E.~D.} \bibnamefont{Rainville}}
  \bibnamefont{and} \bibinfo{author}{\bibfnamefont{P.~E.}
  \bibnamefont{Bedient}}, \emph{\bibinfo{title}{Elementary Differential
  Equations}} (\bibinfo{publisher}{Macmillan}, \bibinfo{year}{1969}),
  p.~\bibinfo{pages}{38}, \bibinfo{edition}{4th} ed.

\bibitem[{\citenamefont{Lucarelli et~al.}(2006)\citenamefont{Lucarelli, Lupke,
  Haugan, Levin, and Barnes}}]{barnes}
\bibinfo{author}{\bibfnamefont{A.}~\bibnamefont{Lucarelli}},
  \bibinfo{author}{\bibfnamefont{G.}~\bibnamefont{Lupke}},
  \bibinfo{author}{\bibfnamefont{T.~J.} \bibnamefont{Haugan}},
  \bibinfo{author}{\bibfnamefont{G.~A.} \bibnamefont{Levin}}, \bibnamefont{and}
  \bibinfo{author}{\bibfnamefont{P.~N.} \bibnamefont{Barnes}},
  \bibinfo{journal}{Supercon. Sci. Technol.} \textbf{\bibinfo{volume}{19}},
  \bibinfo{pages}{667 } (\bibinfo{year}{2006}).

\bibitem[{\citenamefont{Goldacker et~al.}(2006)\citenamefont{Goldacker, Nast,
  Kotzyba, Schlachter, Frank, Ringsdorf, Schmidt, and Komarek}}]{roebel}
\bibinfo{author}{\bibfnamefont{W.}~\bibnamefont{Goldacker}},
  \bibinfo{author}{\bibfnamefont{R.}~\bibnamefont{Nast}},
  \bibinfo{author}{\bibfnamefont{G.}~\bibnamefont{Kotzyba}},
  \bibinfo{author}{\bibfnamefont{S.~I.} \bibnamefont{Schlachter}},
  \bibinfo{author}{\bibfnamefont{A.}~\bibnamefont{Frank}},
  \bibinfo{author}{\bibfnamefont{B.}~\bibnamefont{Ringsdorf}},
  \bibinfo{author}{\bibfnamefont{C.}~\bibnamefont{Schmidt}}, \bibnamefont{and}
  \bibinfo{author}{\bibfnamefont{P.}~\bibnamefont{Komarek}},
  \bibinfo{journal}{J. Phys.: Conf. Ser.} \textbf{\bibinfo{volume}{43}},
  \bibinfo{pages}{901 } (\bibinfo{year}{2006}).

\end{thebibliography}
\end{document}